\documentstyle[prl,aps,epsf,floats]{revtex}
\begin{document}
\twocolumn[
\hsize\textwidth\columnwidth\hsize\csname@twocolumnfalse\endcsname

\draft
\title{Dimensional Crossover in Quantum Antiferromagnets}

\author{Sudip Chakravarty}
\address{Department of Physics and Astronomy\\ University of California Los 
Angeles
\\ Los Angeles, CA 90095-1547}
\date{\today}
\maketitle
\begin{abstract}
The dimensional crossover in a spin-$S$ nearest neighbor
Heisenberg  antiferromagnet is
discussed as it is tuned from  a two-dimensional square
lattice, of lattice spacing $a$, towards a spin chain by varying the
width $L_y$ of a semi-infinite strip $L_x\times L_y$. For integer spins
and arbitrary $L_y$, and for half integer spins with $L_y/a$ an
arbitrary even integer, explicit analytical
expressions for the zero temperature correlation length and the spin gap
are given. For half integer spins and $L_y/a$ an odd inetger, it is
shown that the $c=1$ behavior of the $SU(2)_1$ WZW fixed point is
squeezed out as the width $L_y\to \infty$; here $c$ is the
conformal charge. The results specialized to $S=1/2$ are relevant to
spin-ladder systems.
\end{abstract}
]

\newpage
One-dimensional quantum antiferromagnets have many unusual properties.
For example, nearest-neighbor Heisenberg spin chains with half integer
spins are gapless, but those  with integer spins are generically
gapful\cite{Haldane}. Many properties, such as these, can be understood
from the correspondence of the spin chain to a 
$(1+1)$-dimensional $O(3)$ non-linear $\sigma$-model, augmented by a
term in the action that contributes a phase factor
$e^{i\theta Q}$ to  the
path integral, where $\theta=2\pi S$ ; $S$ is the spin, and $Q$ is an
integer winding number. For
integer spins, $\theta=0\  {\rm mod}\  (2\pi)$, the phase factor is
unity, but for half-integer spins, $\theta=\pi\  {\rm mod}\  (2\pi)$, it
is
$(-1)^Q$. This leads to the crucial difference between the integer and
the half-integer spins. Although the existence of the gap in the
model with
$\theta=0$  can be seen in a variety of ways\cite{theta0}, the
model with  $\theta=\pi$ is more subtle\cite{Affleck}. For
$\theta=\pi$, the short distance behavior is dominated by two weakly
interacting goldstone modes. These are correctly described by the
perturbative renormalization group analysis that is impervious to the
existence of the
$\theta$-term. In the language of conformal field theory, the system is
in the proximity of an infrared unstable fixed point  corresponding to
conformal charge $c=2$. At short distances, there is no distinction
between integer and half integer spins, and the system appears gapful. 
However, on longer scales, the $\theta=\pi$ model flows to
$k=1$ $SU(2)$ Wess-Zumino-Witten model ($SU(2)_1$ WZW), corresponding to
a $c=1$ massless theory. Indeed, all critical theories in two dimensions
must belong to a conformally invariant fixed point.

The  two-dimensional square-lattice, nearest neighbor Heisenberg
antiferromagnet is entirely different.  It is rigorously known that the
ground state is N\'eel ordered for $S\ge 1$\cite{rigorous}, but no such
proofs exist for
$S=1/2$. Nonetheless, the numerical evidence  for an ordered ground
state for $S=1/2$ is strong\cite{Young}. Moreover, the assumption of an
ordered ground state has yielded predictions\cite{CHN} that are
confirmed in neutron scattering experiments\cite{neutron}. Henceforth, I
shall assume that this is also a solved problem, and the correct low
energy theory is a $(2+1)$-dimensional $O(3)$ non-linear
$\sigma$-model, which is in the broken symmetry phase in its ground
state\cite{CHN}. (The possible existence of 
topological  terms were considered and discarded by
a number of authors\cite{Hopf}.) The corresponding elementary
excitations are  weakly interacting Goldstone modes, that is spin waves.
This low-energy, long wavelength model is essentially geometrical and is
almost entirely determined by symmetry\cite{Weinberg} regardless of
whether or not the magnitude of the spin is large. The two needed
phenomenological constants are the spin wave velocity and the spin
stiffness constant. In principle, experiments can determine these
constants and there is no need to rely on a presumed large-$S$ expansion.

It is an intriguing question to ask how a two dimensional  system
would evolve if we began with  a strip, $L_x\times L_y$, and continuously
tuned the system by
varying the width
$L_y$, with $L_x$ kept infinitely large. Would it approach the
one-dimensional limit, and, if so, how would we recover the sensitivity
to the topological angle $\theta$?  This would be  of only theoretical
interest, albeit considerable interest, if it were not for
experiments on spin ladders\cite{ladders} in which
$S=1/2$ systems of varying width are explored. The purpose of the
present paper is an attempt to clarify this crossover and to   
determine  the evolution of the  excitation spectrum.

It has been argued\cite{Khevschenko,Sierra} that spin ladders correspond
to an effective
$(1+1)$-dimensional $O(3)$ non-linear $\sigma$-model with the
$\theta$-parameter given by $\theta=2\pi n_l S$, where $n_l$ is the
number of legs. Thus, for $S=1/2$, the system is gapful for even-leg
ladders and gapless for odd-leg ladders in accord with
experiments\cite{ladders}. There are complimentary theoretical and 
numerical  approaches to
the ladder systems that
are outside the scope of the reasoning in this paper\cite{Rice}.
However, the crossover problem stated above has not
been fully elucidated, although it was anticipated\cite{Sierra}
that, when  approached along the sequence of even-leg ladders, the gap
must collapse exponentially with the increasing width of the system. In
the present paper, I shall show precisely how this happens and derive a
formula that can be checked. At first
sight, the approach to the two-dimensional limit
along the  odd sequence appears to be simple, because they are gapless.
However, this is not so because the two-dimensional problem, which is
insensitive to topology,  is described by two Goldstone modes; 
it cannot be  a straightforward extension of the $c=1$ fixed point of
the $SU(2)_1$ WZW model.

The Euclidean action of the $O(N)$ quantum non-linear $\sigma$-model, in
which one of the dimensions is singled out and is of finite extent $L$,
is
\begin{equation}
{S\over \hbar}={\rho_s^0\over 2\hbar}\int_0^{\beta\hbar}d\tau\int
d^{d-1}x
\int_0^L
dx_1\left[(\partial_{\mu}\hat\Omega)^2+{1\over
c^2}\left(\partial\hat\Omega\over\partial\tau\right)^2\right],
\end{equation}
where the index $\mu$ runs over all the spatial dimensions, $1$
through $d$. The extent of the imaginary time dimension,
$\beta\hbar$, tends to infinity as the temperature $T=1/k_B\beta$ tends
to zero. We shall impose a periodic boundary condition along the
direction 1; the remaining spatial directions will be assumed to be
infinite in extent. The staggered order parameter field of the
antiferromagnet,
$\hat\Omega$, is an
$N$-component unit vector field, which is a function of $(\tau,
x_1,x_2,\ldots, x_d)$; the spin-$S$ antiferromagnet corresponds to
$N=3$. The parameter $\rho_s^0$ is the bare spin stiffness constant at
the spatial cutoff, $\Lambda^{-1}$, of the model, and the parameter $c$
is the spin wave velocity on the same scale. I shall focus on the zero
temperature behavior and report results of a ``Lorentz"
invariant analysis; therefore, the spin wave velocity will not
renormalize. 

The action at $T=0$ is interesting. The extents in all directions,
except $x_1$, are infinite;  along $x_1$ we have a  periodic
boundary condition. The physical problem at $T=0$ is, therefore,
isomorphic to a problem at finite ``temperature", where the
temperature-like variable is
$\varepsilon_L=\hbar c/L$. With proper identifications of the
parameters, it is {\em identical} to that solved  in Ref.
\cite{CHN}. Let us  define two dimensionless bare coupling constants:
\begin{eqnarray}
g_0={\hbar c \Lambda^{d-1}\over \rho_s^0},\nonumber \\
\varepsilon^0_L={\hbar c\Lambda^{d-2}\over L\rho_s^0}.
\end{eqnarray}
The energy like parameter $\varepsilon^0_L$ plays the role of the
dimensionless temperature-like coupling constant in Ref. \cite{CHN},
and $g_0$ is the same as that defined previously. 

The renormalization group equations can be simply  read off from Ref.
\cite{CHN}. The crossover phase diagram, in $d=2$, constructed from these
equations is shown in Fig. 1, merely for orientation.
\begin{figure}[htb]
\centerline{\epsfxsize=2.5 in\epsffile{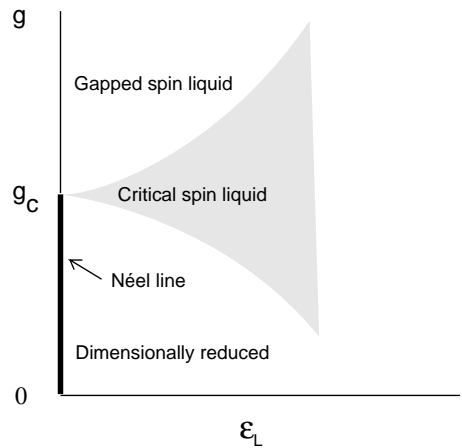}}
\caption{The  crossover phase diagram in $d=2$.}
\label{Fig}
\end{figure}
The three distinct regions had to be renamed, as the present analysis
corresponds to
$T=0$. The regions previously named ``renormalized classical", ``quantum critical", and ``quantum disordered" are now renamed to be ``dimensionally reduced", ``critical spin liquid", and ``gapped spin liquid", respectively.
 With simple
transcriptions, the physical pictures of the crossover boundaries are
the same as before. The analyses of the  gapped and the critical spin
liquid regimes are exceedingly complex and are beyond the scope of the
present paper. 

The region of the phase diagram for which we can make
precise predictions is the dimensionally reduced region. 
In this region, the system is in
the N\'eel ordered state when $L=\infty$, or $\varepsilon_L=0$. When
$L\ne \infty$, the system is equivalent to a dimensionally reduced
effective $(1+1)$-dimensional  model with no long range order. This
can be seen from the renormalization group equations\cite{CHN}. At
first, with increasing length scale, $g$  rapidly decreases, and
$\varepsilon_L$ increases slowly, that is, the system appears more and
more ordered. Subsequently, the growth of $g$ slows down, but 
$\varepsilon_L$ increases more rapidly, thereby breaking up the order at
longer length scales, resulting in the reduction from
$(2+1)$ to $(1+1)$ dimensions. The  effective
coupling constant to be used in the $(1+1)$-dimensional model is easily
calculated to be
\begin{equation}
{1\over \varepsilon_{\rm eff}}={L\over \hbar
c}\rho_s\left[1+{\hbar c\over 2\pi L\rho_s}\ln (\Lambda
L)\right],
\end{equation}
where $\rho_s$ is the fully renormalized macroscopic spin
stiffness constant at $T=0$ of the square-lattice spin-$S$
antiferromagnetic Heisenberg model ($L_x=\infty$, $L_y=\infty$)\cite{foot1}. This definition can be
made more explicit, if we recall that
$\rho_s=JS^2Z_{\rho_s}$, and $\hbar c = 2\sqrt{2}JSaZ_c$, where
$J$ is the exchange constant, $a$ is the lattice spacing; $Z_{\rho_s}$
and $Z_c$ are the renormalization factors\cite{foot3}. We can now write
\begin{equation}
 \varepsilon_{\rm eff}={2\over S}\left[\left({Z_{\rho_s}\over
\sqrt{2}Z_c}\right){L\over a}+{1\over \pi S}\ln (\Lambda L)\right]^{-1}.
\end{equation}
Therefore, for large $(L/a)$, the input bare coupling constant to the
efffective $(1+1)$-dimensional model is greatly reduced from its value
$(2/ S)$ of a spin chain. This is due to increased order at short
distances, concommitant of the quasi two-dimensional nature of the
model with finite width $L$.

For integer spins, and for  even-legged  ladders with half integer
spins, the description of the dimensional crossover is conceptually
complete.  The system is massive and its mass gap should be calculated
with the effective coupling derived above. I shall make more precise
predictions later. The  case of odd-legged ladders with half integer
spins requires further clarification. Because all possible
topological terms were dropped in the $(2+1)$-dimensional model, the
masslessness of the dimensionally reduced system could not be recovered.
Even in the presence of sufficiently strong local N\'eel order, the
proof of the nonexistence of the topological term in
$d=2$ is correct strictly when the number of spins along both $L_x$ and
$L_y$ are even. For odd number of chains along $L_y$, but $L_x=\infty$, a
topological  term $2\pi i S Q$ remains\cite{Hopf}. If we wish, we can
rewrite it as
$2\pi i S Qn_l$ by realizing that the topological
angle is only defined modulo
$2\pi$. It has been never fully explained  why this odd-chain
case is physically irrelevant; in fact, it is not, as we shall see.
Imagine that we include such a term in our action\cite{Marshall}. For
finite
$L$, this should, in principle, render the model massless in the sense of
$c=1$, not in the sense of a Goldstone phase. However, the $c=1$
behavior will be difficult to see when
$L/a$ is large. The reason is that the effective coupling constant
derived in Eq. (4) will be very small, and the perturbative
renormalization group, which is impervious to the topological term, will
be valid up to very long distances, until the coupling constant becomes
of order unity for  the system to crossover to   $c=1$
$SU(2)_1$ WZW model. Recall that the $\theta=\pi$ model in $(1+1)$
dimensions is massless but not conformally invariant; it has
a non-trivial $\beta$-function and an associated mass gap. Thus, the
region in which the $c=1$ behavior is seen is
squeezed out as $L\to \infty$, and all we see is the Goldstone phase;
we shall see that the mass gap vanishes exponentially as $L\to \infty$.
Conversely, when
$(L/a)$ is of order unity, the $c=1$ feature should be visible.

Using Ref. \cite{CHN}, it is  possible to write down by
inspection the expression for the correlation length, $\xi$, in 
our $O(3)$ model defined on a strip for which $L_x=\infty$, $L_y=L$. It
is important to note that the dimensionally reduced effective
$(1+1)$-dimensional model has ``Lorentz" invariance; integrating out
the $L_y$-modes does not destroy the proportionality between the
imaginary time and the $L_x$ directions. There is, therefore, one and
only one correlation length\cite{foot2}. The result for
$\xi$ is
\begin{eqnarray}
\xi&=&\sqrt{32}e^{\pi/2}(2\pi C)\left ({\hbar
c\over 2\pi\rho_s}\right)\exp\left({2\pi\rho_s L\over \hbar
c}\right)\nonumber \\
&&\left[1-A\left({\hbar c\over 2\pi\rho_s
L}\right)+O\left({\hbar c\over 2\pi\rho_s L}\right)^2\right].
\end{eqnarray}
 From strong coupling simulations, the
quantity $2\pi C$ was estimated in Ref.
\cite{CHN} to be between 0.01 and 0.013. This makes the overall
numerical prefactor to be between 0.27 and 0.35. Since then an
asymptotically exact expression has been derived\cite{Hasenfratz}, and
the exact prefactor is known to be
$e/8\approx 0.34$. In addition, the constant which was previously
known to be only of order unity is determined to be $A=1/2$.

It is interesting to rewrite Eq. (5) in terms of  the Josephson
correlation length of the
$L=\infty$  $(2+1)$-dimensional $O(3)$ non-linear
$\sigma$-model\cite{CHN}, which is given by $\xi_J= (\hbar c/\rho_s)$.
This length separates the short distance critical behavior from the long
distance Goldstone behavior. In terms of $\xi_J$, the correlation
length takes the  simple finite-size scaling form:
\begin{equation}
\xi={e\over 8}\left({\xi_J\over 2\pi}\right)e^{2\pi L/
\xi_J}\left(1-{1\over  2}\left(\xi_J/
2\pi L\right)+O\left({\xi_J/ 2\pi L}\right)^2\right),
\end{equation}
For spin-$S$ square-lattice Heisenberg
antiferromagnet, $\xi_J$ is given by\cite{foot3}
\begin{equation}
\xi_J=\left({2\sqrt{2}Z_c\over S Z_{\rho_s}}\right)a.
\end{equation}

Because of ``Lorentz" invariance,  the spin gap, $\Delta$, is simply
$\Delta=\hbar c/\xi$. Note again that the spin wave
velocity does not renormalize and this relation does not have any
corrections ({\em cf.} below). Specializing to
$S=1/2$, we get
\begin{eqnarray}
\left({\xi/ a}\right)_{1\over 2}&=&0.499 e^{0.682(L/ a)}[1-0.734 (a/
L)],
\\
\left({\Delta/ J}\right)_{1\over 2}&=&3.347 e^{-0.682(L/ a)}
\left[1-0.734 (a/ L)\right]^{-1}.
\end{eqnarray}
The field theory analysis presumes that a continuum theory is
applicable in both directions, and therefore 
the expressions in Eqs. (8) and (9) cannot be  accurate for
$(L/a)\sim 1$. Moreover, these expressions, obtained with a
periodic boundary condition, are likely to be different from
those obtained from other boundary
conditions, such as the open boundary condition, especially when
$(L/a)\sim 1$. Nonetheless, if we take $L/a=4$ corresponding to
four-leg ladders, we get
$\left({\Delta/ J}\right)_{1\over 2}=0.268$, and  $\left({\xi/
a}\right)_{1\over 2}=6.23$. For $L/a=6$, we get $\left({\Delta/
J}\right)_{1\over 2}=0.064$, and $\left({\xi/
a}\right)_{1\over 2}=26.2$.

To compare, numerical results, with open boundary condition along $L_y$, are available for $L/a=4$ and 6. According to Ref.
\cite{Scalapino}, we have, for $L/a=4$, $\left({\Delta/
J}\right)_{1\over 2}=0.190$, and $\left({\xi/ a}\right)_{1\over
2}=5-6$. According to Ref. \cite{Greven}, we have, for $L/a=4$,
 $\left({\Delta/ J}\right)_{1\over 2}=0.160$, and $\left({\xi/
a}\right)_{1\over 2}=10.3$. From the same work, we have, for $L/a=6$,
 $\left({\Delta/ J}\right)_{1\over 2}=0.055$. The value for the
correlation length for $L/a=6$ is not given explicitly. However, a
simple extrapolation yields a value close to 30. From Ref. \cite{Frish},
we have, for  $L/a=4$, $\left({\Delta/ J}\right)_{1\over 2}=0.17$;  
for $L/a=6$, $\left({\Delta/ J}\right)_{1\over 2}=0.05$.

 Note  that the gaps should be inversely proportional to the correlation lengths, where the proportionality constant does not renormalize due to ``Lorentz" invariance. We require ``Lorentz" invariance only in the $L_x$-$\tau$ plane. This has little to do with the fact that
the width along  $L_y$ is finite and equal to $L$. This proportionality is automatically satisfied for the analytical expressions given in this paper and should be a good check on the numerical work.

Based on the simple observation that at $T=0$ the spin-$S$ square-lattice
Heisenberg model of finite width  can be mapped on to a
$(2+1)$-dimensional $O(3)$ non-linear
$\sigma$-model with a finite dimension, I have provided a theory for
crossover in spin ladders. Explicit analytical expressions for the
correlation length and the spin gap were obtained by transcribing the
results in Ref. \cite{CHN}. The agreement with numerical calculations is very good considering the difference in the boundary conditions employed.
The analytical expressions show
precisely that the crossover to the two-dimensional limit is approached
exponentially as the the number of legs in the ladder system is
increased. The extension of the present theory to
anisotropic coupling and to finite temperature should be
straightforward. I hope to return to these extensions in the future.

This work was conceived and carried out at the Aspen Center for
Physics. It was supported by a grant from the National Science
Foundation, Grant No. DMR-9531575.


\begin{references}
\bibitem{Haldane}F. D. M. Haldane, Phys. Lett. {\bf 93} A, 464 (983);
Phys. Rev. Lett. {\bf 50}, 1153 (1985); J. Appl. Phys. {\bf 57}, 3359
(1985); I Affleck, Nucl. Phys. B {\bf 257}, 397 (1985).
\bibitem{theta0}A. M. Polyakov, Phys. Lett. {\bf 59} B, 79 (1975); S. H.
Shenker and J. Tobochnik, Phys. Rev. B {\bf 22}, 4462 (1980); C. J.
Hamer, J,. B. Kogut and L. Susskind, Phys. Rev. D {\bf 19}, 3091 (1979);
P. B. Wiegman, Phys. Lett. B {\bf 152}, 209 (1985).
\bibitem{Affleck}I. Affleck and F. D. M. Haldane, Phys. Rev. B {\bf 36},
5291 (1987).
\bibitem{rigorous}T. Kennedy, E. H. Lieb and B. S. Shastry, J. Stat
Phys. {\bf 53}, 1019 (1988); K. Kubo and T. Kishi, Phys. Rev. Lett {\bf
61}, 2585 (1988).
\bibitem{Young} A. P. Young in {\em Les Houches}, session LVI, edited
B. Dou\c{c}cot and J. Zinn-Justin (Elsevier, Amsterdam, 1995) and
references therein.
\bibitem{CHN}S. Chakravarty, B. I. Halperin and D. R. Nelson, Phys.
Rev. Lett. {\bf 60}, 1057 (1988); Phys. Rev. B {\bf 39}, 2344 (1989).
\bibitem{neutron}M. Greven {\em et al.}, Z. Pnys. B {\bf 96}, 465
(1995). For an earlier summary, see S. Chakravarty in {\em High
temperature superconductivity}, edited by K. S. Bedell {\em et al.}
(Addison-Wesley, Redwood City, 1990).
\bibitem{Hopf}See the discussion in E. Fradkin, {\em Field theories
of condensed matter systems} (Addison-Wesley, Redwood City, 1991) and
references therein.
\bibitem{Weinberg}S. Weinberg, Phys. Rev. Lett. {\bf 17}, 616 (1966);
C. G. Callan, S. Coleman, J. Wess and B. Zumino, Phys. Rev. {\bf 177},
2247 (1969).
\bibitem{ladders}D. C. Johnston, J. W. Johnson, D. P. Goshorn, and A.
J. Jacobsen, Phys. Rev. B {35}, 219 (1987). Z. Hiroi, M. Azuma, M.
Takano, and Y. Bando, J. Solid State Chem. {\bf 95}, 230 (1991); M.
Azuma {\em et al.}, Phys. Rev. Lett. {\bf 73}, 3463 (1994).
\bibitem{Khevschenko}D. V. Khveshchenko, Phys. Rev. B {\bf 50}, 380
(1994).
\bibitem{Sierra}G. Sierra, J. Math. Phys. A {\bf 29}, 3299 (1996).
\bibitem{Rice}For a review, see E. Dagotto and T. M. Rice, Science,
{\bf 271}, 618 (1996) and references therein.
\bibitem{foot1}As shown in Ref. \cite{CHN}, a one-loop calculation of
the fluctuations due to the finite size $L$ is sufficient to obtain an
expression correct to  two-loop order in the dimensionally reduced regime.
The matching to strong coupling calculations, combined with a
factor due to conversion of the regularization, then fixes the prefactor.
\bibitem{foot3}In the lowest order spin wave theory, the quantities
$Z_c$ and $Z_{\chi}$ are given by
$Z_c=1+0.158/2S+O(1/2S)^2$, and $Z_{\chi}=1-0.552/(2S)+O(1/2S)^2$. For
$S=1/2$, accurate values are known for $Z_c$ and $Z_{\chi}$:
$Z_c=1.18$, $Z_{\chi}=0.52$ (R. R. P. Singh, Phys. Rev. B {\bf 39}, 9760
(1989); R. R. P. Singh and D. Huse, Phys. Rev. B {\bf 40}, 7247 (1989)).
These values are not very different from those obtained from the lowest
order spin wave perturbation theory, which yields $Z_c=1.15$, and
$Z_{\chi}=0.448$. The quantity $Z_{\rho_s}$ is given by
$Z_{\rho_s}(S)=Z_c^2(S)Z_{\chi}(S)$. For $S=1/2$, I shall use the more
accurate values.
\bibitem{Marshall} For a bipartite lattice, the
Marshall sign condition enforces zero total spin for even number of
spins. If an additional spin is added, the total spin should be $1/2$. This
is true even for the broken symmetry phase, although by appropriately
superposing states (P. W. Anderson, Phys. Rev. {\bf 86}, 694 (1952).) a
direction for the staggered order parameter can be chosen, and
the doublet nature for odd number of spins is irrelevant. However, once 
$L_y$ is finite, the symmetry is restored, and the odd-even alternation
must be allowed. The  way to achieve this in a continuum theory is to
recognize that the topological term is always present, even though its
effect in the broken symmetry state can be ignored.
\bibitem{foot2}In contrast to correlation length, the staggered
susceptibility cannot be so simply obtained from Ref. \cite{CHN}. For
this we need equal time correlation function of the effective
$(1+1)$-dimensional. From Ref.
\cite{CHN}, we can read off trivially only the $q=\pi/a$, $\omega =0$,
susceptibility.
\bibitem{Hasenfratz}P. Hasenfratz and F. Niedermayer, Phys. Lett.
B {\bf 268}, 231 (1991).
\bibitem{Scalapino}S. R. White, R. M. Noack, and D. J. Scalapino,
Phys. Rev. Lett. {\bf 73}, 886 (1994).
\bibitem{Greven}M. Greven, R. J. Birgeneau, and U. -J.  Wiese,
preprint, cond-mat/9605068.
\bibitem{Frish}B. Frischmuth, B. Ammon and M. Troyer, preprint, cond-mat/9601025.
\end{references}
\end{document}